\documentclass[pdflatex,sn-mathphys-ay,iicol]{sn-jnl}

\usepackage{graphicx}%
\usepackage{multirow}%
\usepackage{amsmath,amssymb,amsfonts}%
\usepackage{amsthm}%
\usepackage{mathrsfs}%
\usepackage[title]{appendix}%
\usepackage{xcolor}%
\usepackage{textcomp}%
\usepackage{manyfoot}%
\usepackage{booktabs}%
\usepackage{algorithm}%
\usepackage{algorithmicx}%
\usepackage{algpseudocode}%
\usepackage{listings}%

\usepackage{qcircuit}
\usepackage{physics}

\newcommand{\Id}{{\rm 1\hspace{-0.9mm}l}}

\newcommand{\proj}[1]{\ensuremath{\ketbra{#1}{#1}}}

\theoremstyle{thmstyleone}%
\theoremstyle{thmstyletwo}%

\theoremstyle{thmstylethree}%

\begin{document}

\title{Resource-Efficient Variational Quantum Classifier}


\author*[1,2]{\fnm{Petr} \sur{Pt\'{a}\v{c}ek}}\email{petr.ptacek@vsb.cz}

\author[1]{\fnm{Paulina} \sur{Lewandowska}}\nomail 
\equalcont{These authors contributed equally to this work.}

\author[1]{\fnm{Ryszard} \sur{Kukulski}}\nomail
\equalcont{These authors contributed equally to this work.}

\affil*[1]{\orgdiv{IT4Innovations}, \orgname{VSB~-~Technical University of Ostrava}, \orgaddress{\street{17.~listopadu 2172/15}, \city{Ostrava}, \postcode{708 33}, \country{Czech Republic}}}

\affil[2]{\orgdiv{Faculty of Electrical Engineering and Computer Science}, \orgname{VSB~-~Technical University of Ostrava}, \orgaddress{\street{17.~listopadu 2172/15}, \city{Ostrava}, \postcode{708 33}, \country{Czech Republic}}}


\abstract{
We introduce the unambiguous quantum classifier based on Hamming distance measurements combined with classical post-processing. The proposed approach improves classification performance through a more effective use of ansatz expressivity, while requiring significantly fewer circuit evaluations. Moreover, the method demonstrates enhanced robustness to noise, which is crucial for near-term quantum devices.
 We evaluate the proposed method on a breast cancer classification dataset. 
The unambiguous classifier achieves an average accuracy of 90\%, corresponding to an improvement of 6.9 percentage points over the baseline, while requiring eight times fewer circuit executions per prediction. In the presence of noise, the improvement is reduced to approximately 3.1 percentage points, with the same reduction in execution cost. We substantiate our experimental results with theoretical evidence supporting the practical performance of the approach.
}




\maketitle

\section{Introduction}
Quantum computing is a rapidly evolving field, with growing interest in quantum machine learning (QML) \citep{biamonte2017quantum, QML_Algorithms_applications_and_limitations, schuld2019quantum}, where a central question is whether quantum resources can provide an advantage over classical learning models. Among the proposed approaches, hybrid variational algorithms, such as the variational quantum classifier (VQC) \citep{havlivcek2019supervised, ajibosin2024implementation, yu2021analyzing, zhang2025variational}, have gained particular attention due to their compatibility with noisy intermediate-scale quantum (NISQ) devices \citep{preskill2018quantum}.

In the VQC framework, classical data are embedded into quantum states via a feature map, processed by a parameterized ansatz, and measured to obtain classical outputs. While much prior work has focused on training-related challenges (such as overcoming the issue of barren plateaus \citep{mcclean2018barren}), the prediction stage introduces a fundamental limitation. Due to the inherently probabilistic nature of quantum measurements \citep{tomaz2025quantum}, reliable classification typically requires repeated circuit executions \citep{phalak2023shot}. This repeated measurement process can lead to significant computational overhead, particularly in classification tasks where many predictions are required. Consequently, the cost of inference may become a practical bottleneck, even in the presence of fault-tolerant quantum hardware \citep{preskill1998fault, katabarwa2024early}.


Several approaches have been proposed to mitigate this issue. In particular, \citep{recio2025single} introduced a single-shot quantum classifier, where predictions are inferred from a single circuit execution. However, such methods generally require strong assumptions on data embedding, namely that classes are well separated along nearly orthogonal directions in Hilbert space, which is difficult to achieve in practice. As a result, their performance is highly problem-dependent and typically lacks robustness to noise.

In this work, we address this limitation by introducing the unambiguous quantum classifier, based on a measurement strategy combining the Hamming distance with classical post-processing. We show that this approach reduces the number of required circuit executions while improving classification performance.
We evaluate the method on a breast cancer classification dataset and observe that it achieves an average accuracy of 90\%, improving upon a baseline by 6.9 percentage points, while requiring eight times fewer circuit executions per prediction. Under noisy conditions, the improvement decreases to approximately 3.1 percentage points, while preserving the same reduction in execution cost. These results indicate that the proposed approach provides a favorable trade-off between accuracy and quantum resource requirements. 
 We substantiate our experimental results with theoretical evidence supporting the practical performance of the approach.

This paper is organized as follows.
Section \ref{sec:unambi} presents a description of the three variational quantum classifier approaches compared in this work, including the introduction of the unambiguous quantum classifier.
Section \ref{sec:experiments} is dedicated to experimental results demonstrating the advantage of the unambiguous quantum classifier over the baseline methods.
Section \ref{sec:theory} provides the supporting theoretical analysis.
Finally, concluding remarks are presented in Section \ref{sec:conclusion}.

\section{Variational Quantum Classifiers}\label{sec:unambi}
In this section, we present three types of quantum binary classifier. The first type is a single-qubit general quantum classifier \citep{M1_yin2025application, M1_saxena2022performance, M1_M2_tudisco2025multi}, referred to as the \emph{M1} model. The second type, referred to as the \emph{M2} model, is a quantum classifier based on the Hamming distance measurement \citep{ruan2017quantum, M2_akrom2024variational, M1_M2_tudisco2025multi}, which utilizes classical outcomes obtained from measurements on multiple qubits. Finally, we introduce the unambiguous quantum classifier, referred to as the \emph{M3} model, which combines the \emph{M2} model with a post-processing strategy.

In the following models description, we will assume an odd number of qubits  $k=2n+1, n\in \mathbb{N}$. Each VQC model consists of three stages: data encoding, unitary rotation performed by a parametrized ansatz circuit, and Z-basis measurement returning eigenvalues $(m_j)_{j=0}^{k-1} \subset \{\pm1\}$. The models differ in the way the measurement results are processed, as described below.

\subsection*{Model \emph{M1}} 
In this model, the cost function is determined by the sign of the measured eigenvalue \( m_0 \in \{\pm 1\} \), that is, it depends only on the outcome of the measurement of the first qubit. The predicted class for a single data point is inferred from $T$ measurement outcomes, where $T$ denotes the number of shots. By aggregating the class assignments over all $T$ shots, an empirical class probability is estimated, and the final class label is assigned according to the majority outcome.

\subsection*{Model \emph{M2}} In contrast to the previous model, the cost function depends on the measurements' outcomes from all qubits $m = (m_j)_{j=0}^{k-1}$. The predicted class for a single data point is again determined from $T$ measurement outcomes. For each shot, we obtain a vector of eigenvalues $m$ and the sign of the sum of the measured eigenvalues $\text{sign}(\sum_{j=0}^{k-1}m_j)$ determines the assignment of the class. By aggregating class assignments on all $T$ shots, an empirical class probability is estimated, and the final class label is determined according to the majority outcome.

\subsection*{Model \emph{M3} - Unambiguous quantum classifier}
In the introduced model, the cost function depends on the measurement outcomes of all qubits $m = (m_j)_{j=0}^{k-1}$ and the predicted class for a single data point is again determined from the $T$ measurement outcomes. First, ambiguous measurement outcomes are filtered out. We introduce a model parameter $t= \lceil \frac{k}{2} \rceil, \ldots, k$. An outcome is retained only if the absolute value of the sum of the measured eigenvalues satisfies $|\sum_{j=0}^{k-1} m_j| \ge 2t - k$. For each retained shot, the sign of the sum of the measured eigenvalues $\text{sign}(\sum_{j=0}^{k-1}m_j)$ determines the class assignment. By aggregating the class assignments on all retained valid shots, an empirical class probability is estimated, and the final class label is determined according to the majority outcome.
If no valid shots are obtained, the data point is assigned a class label at random.

\section{Experimental results}\label{sec:experiments}

In this section, we present the results of the experiment on the unambiguous quantum classifier. The implementation is publicly available on GitHub repository at
\href{https://github.com/Weliras/RESOURCE-EFFICIENT-VARIATIONAL-QUANTUM-CLASSIFIER}{https://github.com/Weliras/RESOURCE-EFFICIENT-VARIATIONAL-QUANTUM-CLASSIFIER}.
All of  our experiments are implemented in the Python PennyLane framework version 0.40.0 \citep{bergholm2018pennylane}, which is commonly used in the field of QML. We utilized the C++ \emph{lightning.qubit} performance-optimized state-vector simulator for training our QML models, while the \emph{default.mixed} simulator was used for testing under both noiseless and noisy conditions. The implementation is carried out in a series of \emph{Jupyter Notebooks}, which provide an effective environment for experimentation and data visualization.

To compare and evaluate our models against results reported in previous studies, we used a publicly available medical dataset. This dataset was chosen because it represents a common use-case that supports the timely diagnosis of severe health conditions, highlighting the practical relevance and potential impact of the proposed model. Specifically, the chosen dataset is suitable for machine learning applications that aim to predict the presence of breast cancer. The selection of this dataset is further motivated by its suitability for a problem formulated as a binary classification task, which aligns well with the implementation of VQC models.
The dataset used in this study is available at  \citep{dataset}. 

We follow the preprocessing strategy described in \citep{ajibosin2024implementation}. Briefly, we isolate the target column, scale the original input features to a normalized range using sklearn’s \emph{StandardScaler}, and apply Principal Component Analysis (PCA) to select the most relevant features while reducing the dimensionality of the input data. These components represent linear combinations of the original features that capture the greatest variance within the dataset. The resulting PCA components are scaled in the range $[0,1]$ using sklearn’s \emph{MinMaxScaler}. Since the original datasets contain many features and we use an angle embedding method which requires one qubit per encoded feature, we limit the number of features accordingly. The train and test datasets are split into an 80:20 ratio.

\subsection{Circuit's architecture }

In all subsequent circuits and models, the number of qubits used will be denoted as $k$. The number $k$ is set equal to the number of input data features $x_i = (x_{i}^{(0)}, \ldots, x_{i}^{(k-1)})$, where $x_i$ is the $i$-th data point of the selected dataset.
In the quantum circuits of our models, we utilize a layered ZZ-entangling feature map \citep{havlicek2019supervised, tjandra2023evolutionary} $U_{FM}(x_i)$ as we can see in Fig.~\ref{fig:circuit_fm}. 

\begin{figure*}[htbp!]
    \centering
    \resizebox{0.85\linewidth}{!}{$
    \Qcircuit @C=0.4em @R=1.2em {
     & & \hspace{38.0em}\text{Repeated $l_{FM}$ times} & \span & \span & \span & \span & \span & \span & \span & \span & & \\
     & \qw & \gate{H} & \gate{P(2 * x_i^{(0)})} & \targ & \qw & \targ & \cds{4}{\vdots} & \qw & \qw & \qw & \qw & \qw \\
     & \qw & \gate{H} & \gate{P(2 * x_i^{(1)})} & \ctrl{-1} & \gate{P(2 *(\pi - x_i^{(0)}) * (\pi - x_i^{(1)}))} & \ctrl{-1} & \qw  & \qw & \qw & \qw & \qw & \qw \\
     & & & & & & & & & & & &  \rstick{U_{FM}(x_i)\ket{0}^{\otimes k}} \\
     & \qw & \gate{H} & \gate{P(2 * x_i^{(k-2)})} & \qw & \qw & \qw & \qw & \targ & \qw & \targ & \qw & \qw \\
     & \qw & \gate{H} & \gate{P(2 * x_i^{(k-1)})} & \qw & \qw & \qw & \qw & \ctrl{-1} & \gate{P(2 *(\pi - x_i^{(k-2)}) * (\pi - x_i^{(k-1)}))} & \ctrl{-1} & \qw & \qw \inputgroupv{2}{6}{0.8em}{4.8em}{\hspace{-2.5em}\ket{0}^{\otimes k}} \gategroup{2}{3}{6}{11}{1.2em}{--} 
    }
    $}
    \caption{Quantum circuit of ZZ-entangling feature map $U_{FM}(x_i)$ for data point $x_i$ of $k$ data features. At the input, we consider a tensor product state of $k$ qubits, each set in an initial $\ket{0}$ quantum state. After initializing each qubit in an equal superposition state, we apply single-qubit phase rotations $P(\varphi)$ with rotation angle $\varphi$, followed by two-qubit entangling gates (CNOT). This procedure yields the feature-mapped quantum state $U_{FM}(x_i)\ket{0}^{\otimes k}$ for data point $x_i$.}

    \label{fig:circuit_fm}
\end{figure*} 

The feature map $U_{FM}(x_i)$ itself consists of repeating layers $l_{FM}$, allowing a sufficiently deep embedding to effectively separate data points belonging to different classes. A study by Tomono and Natsubori \citep{tomono2023shipping} has highlighted the effectiveness of using ZZ feature maps, demonstrating competitive performance scores in classification tasks compared to other types of feature maps, namely Y, Z-ZZ.
The $U_{FM}(x_i)$ feature map applies controlled-NOT ($\mathrm{CNOT}$) entangling gates between neighboring qubits interleaved with single-qubit rotations, allowing for effective nonlinear mappings of classical features into the Hilbert space. 

We employ a layered trainable ansatz with reverse linear entangling, commonly used in variational algorithms for classification or chemistry applications. This ansatz, a heuristic trial wave function, is implemented in Qiskit as the \emph{RealAmplitudes} class \citep{vedavyasa2024classification}. The variational ansatz consists of $l_{A}$ alternating layers of Y-rotations and $\mathrm{CNOT}$ entangling gates. Chandrasekhar et al. \citep{chandrasekhar2025adapting} have demonstrated the efficiency of this ansatz combined with the ZZ feature map, motivating our choice of the same ansatz. We can see the structure of used ansatz in Fig.~\ref{fig:circuit_ansatz}.

\begin{figure*}[htbp!]
    \centering
    \resizebox{0.65\linewidth}{!}{$
    \Qcircuit @C=0.8em @R=1.2em {
     & \span & \hspace{5.0em}\text{Repeated $l_{A}$ times, $l =$ current layer} & \span & \span & \span & \span & \span & & \\
     & \gate{R_Y(\theta_{0,0})} & \gate{R_Y(x_i^{(0)})} & \qw & \qw & \cds{4}{\vdots} & \targ & \gate{R_Y(\theta_{l,0})} & \qw & \qw \\
     & \gate{R_Y(\theta_{0,1})} & \gate{R_Y(x_i^{(1)})} & \qw & \qw & \qw  & \ctrl{-1} & \gate{R_Y(\theta_{l,1})} & \qw  & \qw \\
     & & & & & & & & &  &  &\rstick{\hspace{-1.5em}U_{A}(x_i, \theta)U_{FM}(x_i)\ket{0}^{\otimes k}} \\
     & \gate{R_Y(\theta_{0,k-2})} & \gate{R_Y(x_i^{(k-2)})} & \targ  & \qw & \qw & \qw & \gate{R_Y(\theta_{l,k-2})} & \qw & \qw \\
     & \gate{R_Y(\theta_{0,k-1})} & \gate{R_Y(x_i^{(k-1)})} & \ctrl{-1} & \qw & \qw & \qw & \gate{R_Y(\theta_{l,k-1})} & \qw & \qw \inputgroupv{2}{6}{0.8em}{4.8em}{\hspace{-4.5em}U_{FM}(x_i)\ket{0}^{\otimes k}} \gategroup{2}{3}{6}{8}{1.0em}{--} 
    }
    $}
    \caption{Parametrized circuit (ansatz) $U_A(x_i, \theta)$ with reverse-entangling and data re-uploading for trainable parameters $\theta_{l,j}$ for current layer $l$ on $j$-th qubit. This ansatz implements data re-uploading  \citep{perez2020data}, introducing the data point $x_i$ to each layer. At the input, we consider an entangled state $U_{FM}(x_i) \ket{0}^{\otimes k}$. At the output we already have $U_A(x_i, \theta)U_{FM}(x_i) \ket{0}^{\otimes k}$, with processed quantum state by ansatz.}
    \label{fig:circuit_ansatz}
\end{figure*}

\subsection{VQC classifier and postprocessing methods}

The models employ a hybrid parameterized quantum circuit composed of $U_{FM}(x_i)$ feature map and $U_A(x_i, \theta)$ ansatz. The resulting circuit $U_{VQC}(x_i, \theta)$ is depicted on Fig.~\ref{fig:circuit_qnn}. 

\begin{figure*}[htbp!]
    \centering
    $\Qcircuit @C=2.0em @R=1.2em {
             & \qw & \multigate{2}{U_{FM}(x_i)} & \qw & \multigate{2}{U_A(x_i, \theta)} & \meter \\
             & \push{\vdots} & \nghost{U_{FM}(x_i)} & & \nghost{U_A(x_i, \theta)} & \push{\vdots} \\
             & \qw & \ghost{U_{FM}(x_i)} & \qw & \ghost{U_A(x_i, \theta)} & \meter \inputgroupv{1}{3}{0.8em}{2.4em}{\hspace{-0.2em}\ket{0}^{\otimes k}}
    }
    $
    \caption{VQC circuit $U_{VQC}(x_i, \theta)$ composed of layered feature map $U_{FM}(x_i)$ with $l_{FM}$ layers and ansatz $U_A(x_i, \theta)$ with $l_{A}$ layers. Input data point is depicted as $x_i$ and trainable weights as $\theta$. The output is obtained by sampling Pauli-Z operator on each qubit.}
    \label{fig:circuit_qnn}
\end{figure*}

The output of the model can be interpreted in two ways \citep{pennylanesample, pennylaneexpval}. First, each $j$-th qubit can be measured in the computational Z-basis using the Pauli-Z operator, which yields sampled eigenvalues $m_j \in \{\pm1\}$. Alternatively, we can directly estimate the expectation value of the Pauli-Z operator on each qubit from repeated measurement samples. For the considered models, both measurement strategies are valid and, after appropriate postprocessing, provide equivalent information for classification. Estimating directly the expectation values offers a practical advantage within the  PennyLane framework, as it directly yields a smoother cost function that is naively compatible with gradient-based optimization methods. Although we experimentally evaluated both approaches, we ultimately chose to sample the eigenvalues of the Pauli-Z operator on each qubit, as this provides greater control over the information that can be extracted from the classical outcomes.

\subsection{Training description} Having described the complete data preprocessing pipeline and model setup, we now turn to the description of the model training procedure. The trainable parameters of each model were optimized individually, using the corresponding postprocessing strategy. We explore a variety of different classical optimizers such as PennyLane native ones: \emph{AdagradOptimizer}, \emph{AdamOptimizer}, \emph{RMSPropOptimizer}, \emph{NesterovMomentumOptimizer} but also \emph{SPSAOptimizer}. In the earliest experiments, we investigate the performance of non-native PennyLane optimizers, including Nelder–Mead and Particle Swarm Optimization, and confirm that they are also applicable. Ultimately, we chose to use the \emph{SPSAOptimizer} \citep{spall1998implementation}, as it is particularly well-suited for noisy environments and requires only two cost function evaluations per optimization step, regardless of the dimension of the input vector. We use \emph{Binary Cross Entropy (Log Loss)} as the cost function. The number of training epochs is capped at 200, with early stopping applied once the cost function has converged. Within each epoch, the cost function is computed in batches of 8 input train data points. The entire training process is logged for subsequent analysis.

By individually training each model using the same optimization settings, we ensure a fair comparison of classification performance across different postprocessing strategies. Each model is evaluated and compared under both noiseless and noisy conditions. For each evaluation, 25 classification runs with trained weights are performed on the full test dataset to collect statistics on evaluation metrics from each run, including the required average count of quantum circuit executions-per-prediction, accuracy and precision. The classification performed by the \emph{unambiguous quantum classifier} (model \emph{M3}) uses acceptance threshold set to $t=3$ for $k=3$ qubits and to $t=4$ for $k=5$ qubits.

\subsection{Noise model description}
We model realistic quantum noise by explicitly introducing standard quantum error channels. The noise is introduced using the PennyLane transform function applied to a quantum circuit, ensuring that the noise channels were inserted immediately after each quantum operation. Specifically, for each qubit on which any operation is done, we sequentially applied: 
\begin{enumerate}
    \item Depolarizing noise (in PennyLane: \emph{DepolarizingChannel} class) with probability $0.02$, modeling the loss of information due to random flips in the quantum state.
    \item Amplitude damping (in PennyLane: \emph{AmplitudeDamping} class) with probability $0.05$, capturing energy relaxation processes in which the qubit state decays from the excited state $\ket{1}$ to the ground state $\ket{0}$.
    \item Phase damping (in PennyLane: \emph{PhaseDamping} class) with probability $0.03$, representing decoherence that reduces the off-diagonal elements of the quantum state density matrix without energy loss.
\end{enumerate}
Formally, if $\mathcal{E}_\mathrm{depol}$, $\mathcal{E}_\mathrm{amp}$, and $\mathcal{E}_\mathrm{phase}$ denote the respective noise channels, for each operation $O$ acting on the qubit $j$, the noisy operation sequence becomes \begin{equation}
O_j \;\; \longrightarrow \;\; 
\mathcal{E}_\mathrm{depol}^{(j)} \;\; \longrightarrow \;\;
\mathcal{E}_\mathrm{amp}^{(j)} \;\; \longrightarrow \;\;
\mathcal{E}_\mathrm{phase}^{(j)}.
\end{equation}

\subsection{Results}

In Fig.~\ref{fig:res_classification_data=C_qubits=3} and Fig.~\ref{fig:res_classification_data=C_qubits=5}, we compare the models \emph{M1}, \emph{M2}, and \emph{M3} with respect to the validation metrics evaluated on the test dataset. The validation metrics includes accuracy, precision, and the average count of quantum circuit executions required to generate a prediction for a single test data point. The plots correspond to experiments conducted with 3 qubits and 5 qubits, respectively. The figures present statistics obtained from 25 classification runs per model in both noiseless and noisy scenarios. The first row of subfigures compares the classification performance of the models using the accuracy metric, while the second row presents the precision scores. The third row, whereas, compares the average count of circuit executions each model requires to make a near-deterministic prediction. The three columns represent the results for individual models, respectively. The solid curves in the third row subfigures represent results obtained in a noiseless environment, whereas the dashed curves correspond to results obtained under noisy conditions.

\subsubsection{3-qubit experiment}
\begin{figure*}[htbp!] 
    \centering
    \includegraphics[width=0.95\textwidth]{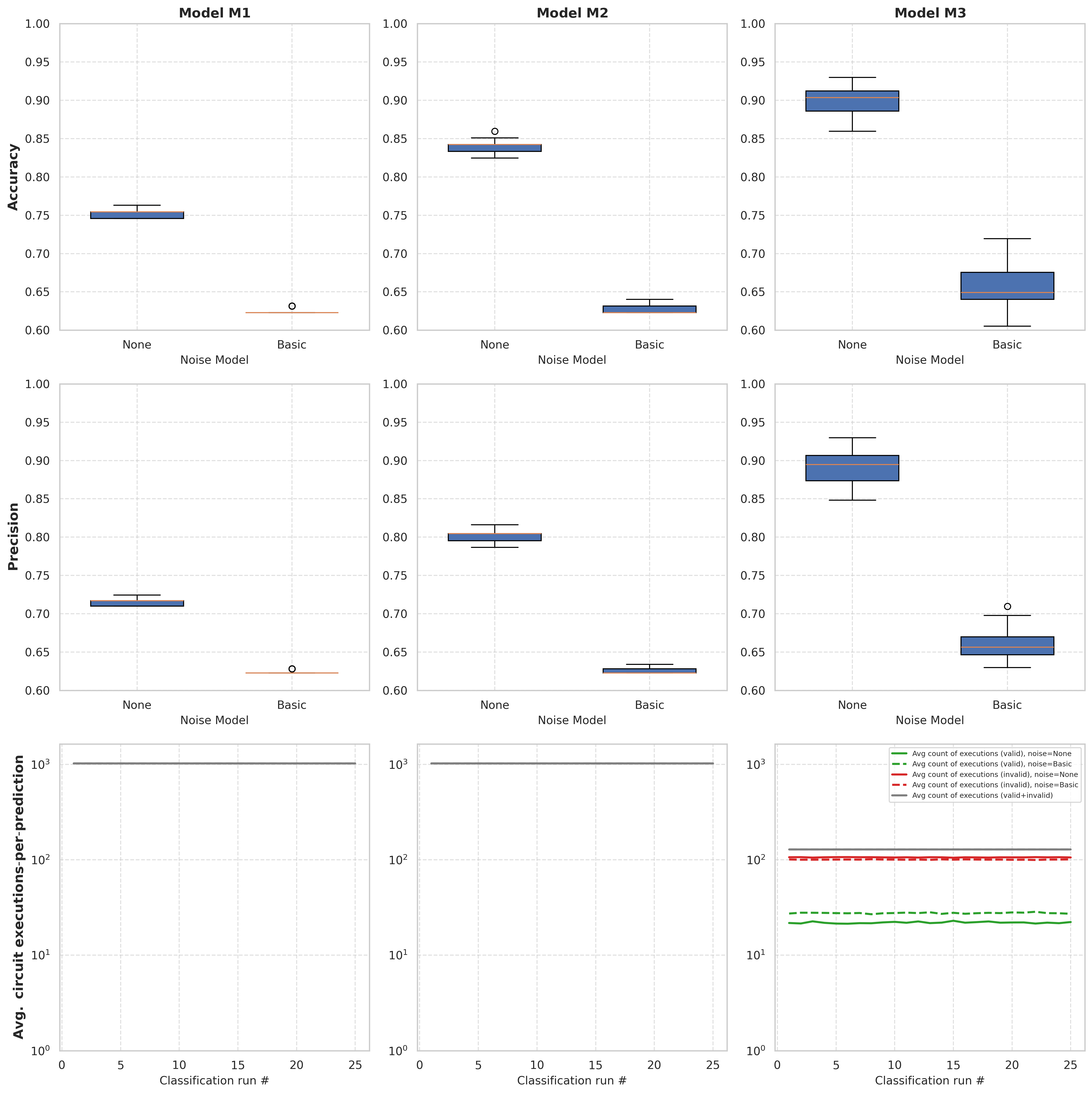}
    \caption{Comparison of classification performance for three-qubits experiment (first and second row subfigures; y-axis in linear scale) and the number of circuit executions per prediction (third row subfigures; y-axis in logarithmic scale) across 25 classification runs for each model, using a circuit with 3 qubits, 2 ansatz layers, and 1 feature map layer.}
    \label{fig:res_classification_data=C_qubits=3}
\end{figure*}

From Fig.~\ref{fig:res_classification_data=C_qubits=3}, we observe that in the noiseless setup the \emph{unambiguous quantum classifier} (model \emph{M3}) achieves the highest average accuracy of 90.00\%. Model \emph{M2} achieves an average accuracy of 83.92\%, while model \emph{M1} achieves the lowest accuracy of 75.15\%.  Model \emph{M1} performs worse than \emph{M3}, confirming our theoretical prediction that the global cost function performs better, because more data is obtained from multiple measurement results. Models \emph{M1} and \emph{M2} were executed 1024 for 1 prediction, while model \emph{M3} was executed only 128 times, on average making 1 prediction from 21.98 valid executions. This implies that models \emph{M1} and \emph{M2} require a total of 116{,}736 circuit executions to generate predictions for the entire test set of 114 samples, while model \emph{M3} requires only approximately 14{,}592 executions. In summary, the proposed \emph{unambiguous quantum classifier} requires $8\times$ fewer circuit executions-per-prediction than the baseline models, while improving accuracy by 13\% compared to the best-performing baseline model \emph{M2}.

In the noisy setting, the accuracy of the best-performing model, \emph{M3}, drops to 65.72\%, representing a decrease of nearly 24.28\% due to noise. Nevertheless, the proposed model still achieves the highest accuracy in noisy setup compared to the baseline models. This significant reduction can be attributed to the use of highly noisy channels with large error probabilities, as well as the absence of standard quantum error correction techniques.

\subsubsection{5-qubit experiment}
\begin{figure*}[htbp!] 
    \centering
    \includegraphics[width=0.95\textwidth]{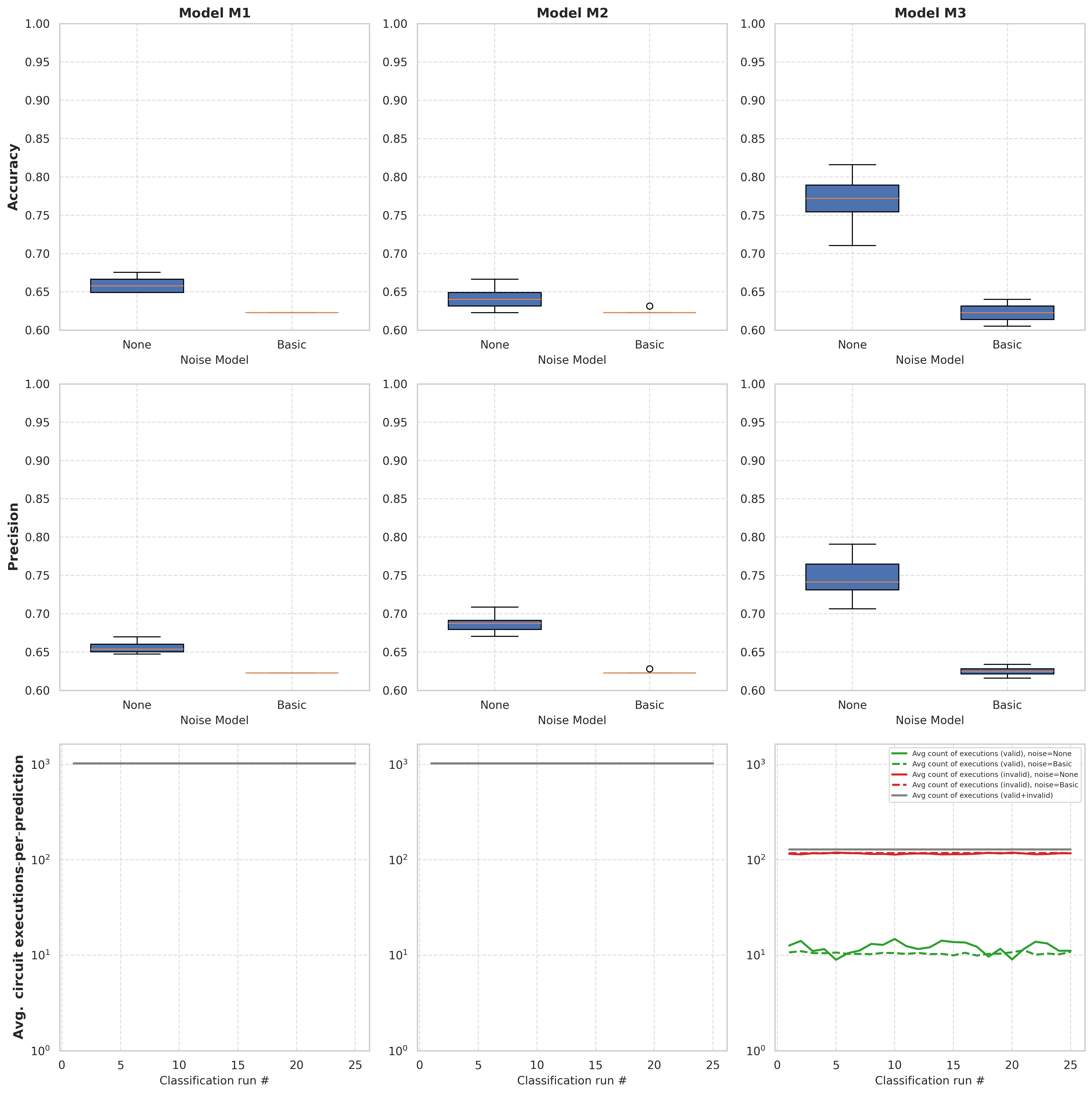}
    \caption{Comparison of classification performance for five-qubit experiment (first and second row subfigures; y-axis in linear scale) and the number of circuit executions per prediction (third row subfigures; y-axis in logarithmic scale) across 25 classification runs for each model, using a circuit with 5 qubits, 2 ansatz layers, and 1 feature map layer.}
    \label{fig:res_classification_data=C_qubits=5}
\end{figure*}

When examining 5-qubit experiment represented in Fig.~\ref{fig:res_classification_data=C_qubits=5}, we observe that model \emph{M3} performs worse on 5 qubits than on 3 qubits. However, it still achieves the highest performance among all models, with an average accuracy of 76.84\% in the noiseless setup and 62.31\% under noisy conditions. Table \ref{tab:model_performance_C} presents the results in tabular form.

\begin{table*}[htbp]
\centering
\begin{tabular}{lccccccc}
\hline
\textbf{Model} & \textbf{Qubits} & \textbf{Noise} & \textbf{Avg. executions (valid)} & \textbf{ACC} \\
\hline
M1 & 3 & NO & 1024 & 0.751579 \\
M2 & 3 & NO & 1024 & 0.839298  \\
M3 & 3 & NO & 128 (21.98) & \textbf{0.900000}  \\
M1 & 3 & YES & 1024 & 0.624211  \\
M2 & 3 & YES & 1024 & 0.626316  \\
M3 & 3 & YES & 128 (27.65) & \textbf{0.657193}  \\
M1 & 5 & NO & 1024 & 0.658596  \\
M2 & 5 & NO & 1024 & 0.641404 \\
M3 & 5 & NO & 128 (12.08) & \textbf{0.768421} \\
M1 & 5 & YES & 1024 & 0.622807  \\
M2 & 5 & YES & 1024 & 0.623158  \\
M3 & 5 & YES & 128 (10.47) & \textbf{0.623158}  \\
\hline
\end{tabular}
\caption{Comparison of model performance with respect to average number of executions-per-prediction and test dataset average accuracy (ACC) score.}
\label{tab:model_performance_C}
\end{table*}

We compared our results with those reported in \citep{ajibosin2024implementation} and found that our models achieve even better performance than their VQC model, referred to as QNN. The comparison was conducted on the same dataset, using the identical feature map and a similar ansatz configuration. The largest improvement observed corresponds to a maximum accuracy increase of 13\%. This improvement can likely be attributed to the use of the proposed post-processing strategy, data re-uploading technique, as well as experimentation with different optimizers, their hyperparameters, and batch sizes. 

\subsubsection{Training history model}
\begin{figure*}[htbp!] 
    \centering
    \includegraphics[width=0.8\textwidth]{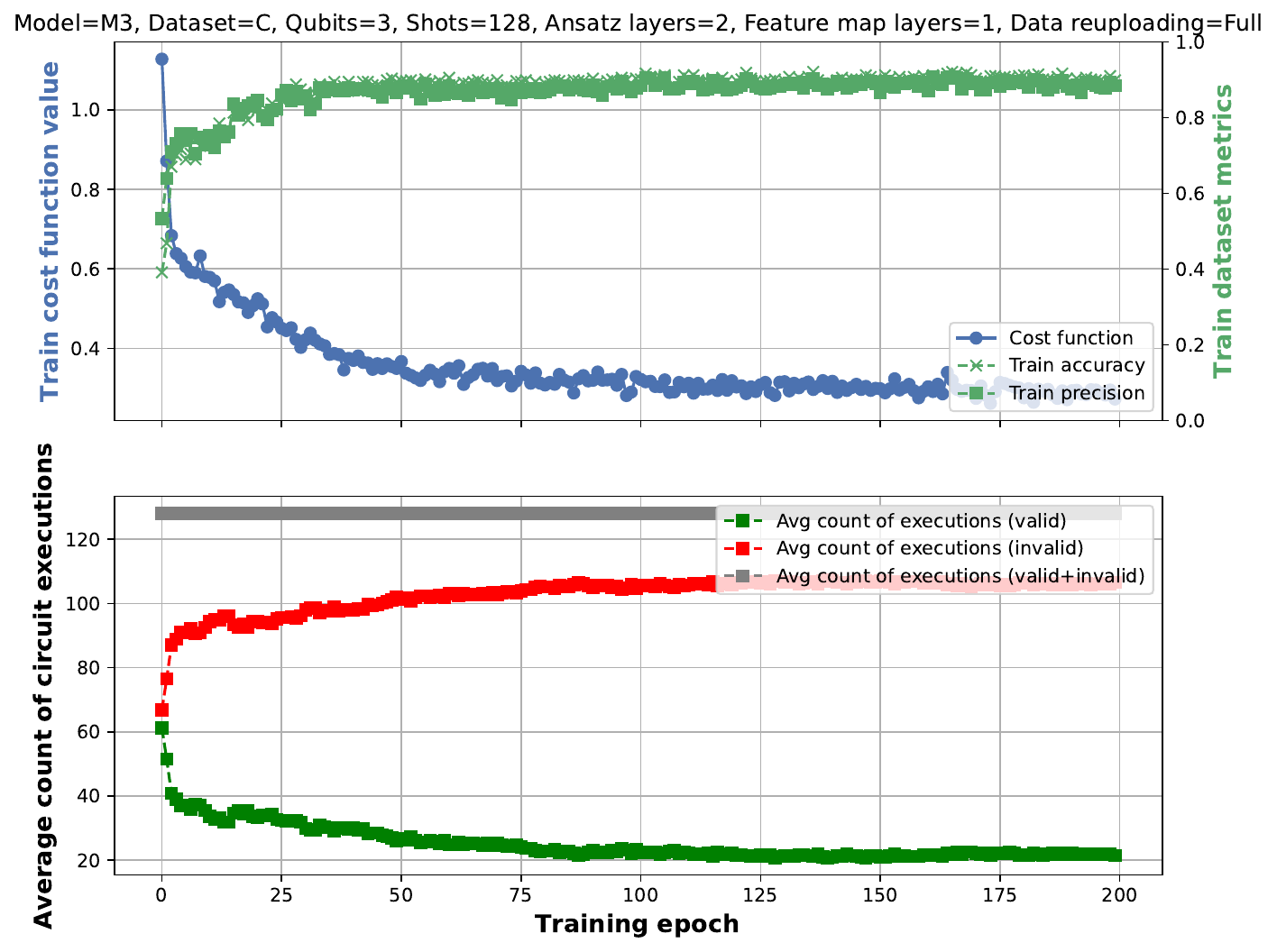} 
    \caption{Training history and average count of valid/invalid/total executions-per-prediction in each training epoch for model \emph{M3} on 3 qubits with 2 ansatz layers, 1 feature map layer using 128 shots.}
    \label{fig:res_training_history_model=M3_data=C_qubits=3}
\end{figure*}

\begin{figure*}[htbp!] 
    \centering
    \includegraphics[width=0.8\textwidth]{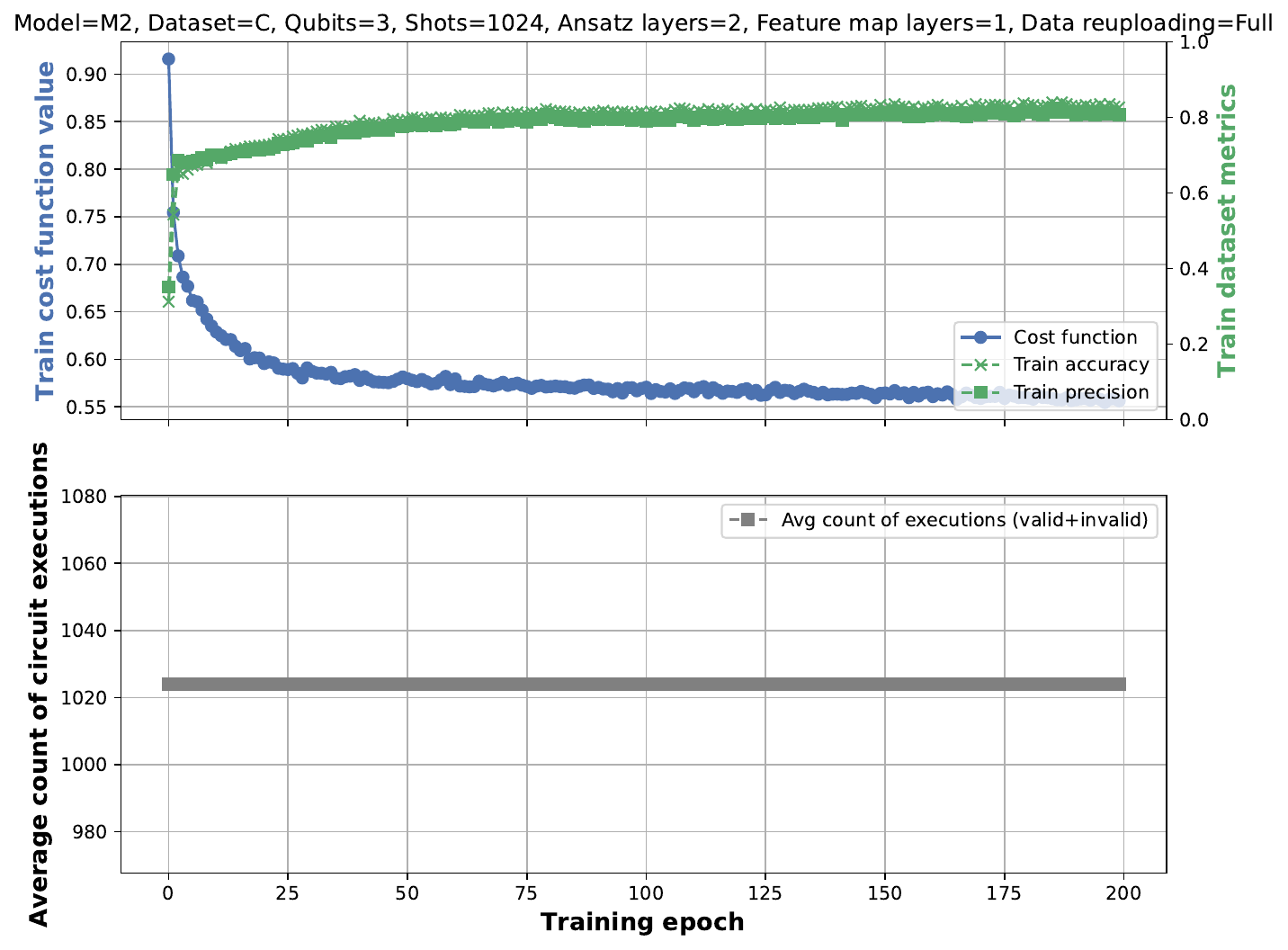} 
    \caption{Training history and average count of total executions-per-prediction in each epoch for model \emph{M2} on 3 qubits, with 2 ansatz layers, 1 feature map layer using 1024 shots.}
    \label{fig:res_training_history_model=M2_data=C_qubits=3}
\end{figure*}

We have trained our \emph{unambiguous quantum classifier} to evaluate how well it could optimize the weights under the constraint of using as few executions as possible. We compared it with the training that we normally employ on model \emph{M2} and the comparison can be seen in Fig.~\ref{fig:res_training_history_model=M3_data=C_qubits=3} and Fig.~\ref{fig:res_training_history_model=M2_data=C_qubits=3}. For training, we used the PennyLane \emph{SPSAOptimizer} \citep{spall1998implementation}, as training model \emph{M3} involves significant noise due to the inherent randomness of quantum measurements and the constraint on the number of executions. We can see that model \emph{M3} is capable of training itself; however, this approach introduces noise into the training process, causing the cost function to oscillate. Despite this, the optimizer successfully minimizes the cost and achieves convergence.

\section{Theoretical analysis of results}\label{sec:theory}

The experimental results provide premise that model \emph{M3} achieves better results than \emph{M1} or \emph{M2} while requiring less quantum resources. In this Section we would like to support this claim by semi-theoretical analysis that will be focused on Accuracy comparison.  

Let $x_i$ be given $i$-th input data point which is encoded into quantum circuit and it is assumed to be of the class ``0'' (an alternative class will be enumerated by ``1''). After the training stage in model \emph{M1} the ansatz transforms $x_i$ into quantum state $\rho_1$ on $k$ qubits, which will be measured in $Z$ basis. Let $p_0^{M1} = \tr\left(\rho_1 (\proj{0} \otimes \Id) \right)$ be the probability for \emph{M1} that $x_i$ will be classified as class ``0'' in single-shot scenario ($T=1$). Similarly, let $\rho_2$ and $\rho_3$ be encoded states for $x_i$ in models \emph{M2} and \emph{M3}, respectively. For the second model we define $p_0^{M2} = \tr\left(\rho_2 \Pi_0 \right)$, where $\Pi_0 = \sum_{b: \sum b_i <k/2} \proj{b}$ is the projector onto the class ``0'' subspace which consists of bit-strings $b = (b_0,\ldots, b_{k-1}) \in \{0,1\}^k$ that number of ``zeros'' is greater than ``ones'', $b_0 +\ldots+b_{k-1} <k/2$. As measurement projectors $\proj{0} \otimes \Id$ and $\Pi_0$ are unitarily similar, it should be possible for complex enough parametrized circuit to rotate the data appropriately during training stage. Therefore, we can fairly assume that training quality of models \emph{M1} and \emph{M2} is equal, $p_0^{M1} = p_0^{M2} > 0.5$. For the model \emph{M3} we choose the parameter $t > \lceil \frac{k}{2} \rceil$ ($t=\lceil \frac{k}{2} \rceil$ case corresponds to the model \emph{M2}). The state $\rho_3$ is assumed to be with the probability $p_0^{M2}$ inside the projector $\Pi_0$, $\tr\left( \rho_3 \Pi_0 \right) = p_0^{M2}$. However, inside the subspace $\Pi_0$ (as well as $\Id - \Pi_0$) the ansatz can rotate further the state to increase the accuracy for model \emph{M3}. Let us define a projector $\Sigma_0 = \sum_{b: \sum b_i \le k-t} \proj{b}$, which consists of bit-strings that accept class ``0'' in single-shot scenario. Additionally, let $\Sigma_1$ will be defined similarly, but for the class ``1'' and $\Sigma_? = \Id - \Sigma_0 - \Sigma_1$ denote the case, where the measurement output is rejected.

To compare accuracy for arbitrary $T$ we denote $p_0 = p_0^{M2}, p_1=1-p_0$ and $q_0 = \tr(\rho_3 \Sigma_0), q_1 = \tr(\rho_3 \Sigma_1), q_? = \tr(\rho_3 \Sigma_?) $. For each model \emph{M} we define the probability of successful classification in $T$ shots as $P_{M,T}$. Applying Stirling's approximation formula one can show~\citep{ash2012information} for model \emph{M2} (being equal to \emph{M1}) that $P_{M2,T} \le 1 - \frac{1}{\sqrt{2T}} \exp(-T D(\frac12 || p_0))=1 - \frac{1}{\sqrt{2T}} \left( 2\sqrt{p_0p_1} \right)^T$. On the other hand, the Chernoff bound for model \emph{M3} yields following lower-bound $P_{M3,T} \ge 1- \left(2\sqrt{q_0q_1}+q_?\right)^T$. Hence, for large enough $T$ model \emph{M3} gains advantage over \emph{M1} and \emph{M2} as long as
\begin{equation}\label{eq-th}
 2\sqrt{p_0p_1} > 2\sqrt{q_0q_1}+ q_?.
\end{equation}
There are two extreme scenario during the training phase for \emph{M3}. If the classifier  do not rotate bit-strings further inside $\Pi_0$ then $p_0/q_0 = p_1/q_1$ and Eq~\eqref{eq-th} does not hold. However, the best classifier will rotate information in such a way that maximizes success $q_0 \to p_0$ and reduces error $q_1 \to 0$. Then Eq.~\eqref{eq-th} holds and \emph{M3} provides advantage. 

The real classifier does not fall into any of these categories but interpolates them. That leaves the window for results improvement as observed in Fig.~\ref{fig:res_classification_data=C_qubits=3} and Fig.~\ref{fig:res_classification_data=C_qubits=5}. The choice of parameter $t$ plays also important role here. If $t$ is nearby $k$ it will be easier to achieve $q_1 \simeq 0$ but harder for $q_0 \simeq p_0$ and when $t$ is nearby $k/2$ then reversely, $q_0 \simeq p_0$ is easier to achieve for the price of $q_? \simeq 0$. 

Finally, it is worth to empathize that $\epsilon$ strong errors linearly interfere with the accuracy for model \emph{M1} - any bit-flip on first qubit changes the class. For \emph{M2} the interference is also linear, but the factor depends on the distribution of given bit-strings encoding any input state. In the worst-case scenario model \emph{M2} shall behave slightly worse than \emph{M1} in noisy computation. Nevertheless, we can observe natural quadratic error suppression for model \emph{M3} as all non-decisive cases falling into subspace $\Sigma_?$ provide a non-trivial barrier separating both classes. 

\section{Conclusion and discussion}\label{sec:conclusion}
In this work, we introduced the \emph{unambiguous quantum classifier}, a measurement strategy designed to improve computational accuracy while mitigating the stochastic overhead inherent in quantum classification. The proposed approach reduces the number of required quantum circuit executions without modifying the underlying variational ansatz. Additionally, we showed that the unambiguous quantum classifier exhibits enhanced noise resistance, which is crucial for near-term quantum devices.

Experimental results on a publicly available dataset show that, in a 3-qubit noiseless setting, the proposed classifier achieves an average accuracy of 90\%, outperforming the best-performing baseline by 6.9 percentage points, while requiring \(8\times\) fewer circuit executions per prediction. In the presence of noise, the improvement remains moderate (approximately 3.1 percentage points), while preserving the same reduction in computational cost.

The observed performance gains can be attributed to the combination of post-processing based on filtering ambiguous measurement outcomes and the use of data re-uploading, which improves training convergence. Furthermore, the model remains trainable under realistic conditions using stochastic optimizers such as SPSA.

Comparison with recent studies, including \citep{ajibosin2024implementation}, shows that the proposed approach achieves consistently higher accuracy under comparable settings, with a maximal improvement of up to 13\%.

Overall, these results demonstrate that appropriate post-processing strategies can significantly reduce measurement-induced stochastic overhead while improving classification performance. This provides a practical pathway toward more resource-efficient variational quantum classifiers suitable for near-term quantum hardware.

Future work will focus on extending the approach to larger systems and more complex datasets, investigating adaptive acceptance thresholds, and incorporating noise-aware training and error mitigation techniques.

\section*{Declarations}

\textbf{Funding} 
PP is supported by the Ministry of Education, Youth and Sports of the Czech Republic through the e-INFRA CZ (ID:90254) and by a Grant of SGS No. SP2026/050, VŠB - Technical University of Ostrava, Czech Republic.

PL is supported by the Ministry of Education, Youth and Sports of the Czech Republic through the e-INFRA CZ (ID:90254), with the financial support of the European Union under the REFRESH – Research Excellence For REgionSustainability and High-tech Industries project number \verb|CZ.10.03.01/00/22_003/0000048| via the Operational Programme Just Transition.  

RK is supported by the European Union under the Quantum error correction codes enhanced by reinforcement learning dedicated for the Ising model-based optimization, contract nr. 01906/2025/RRC via the Operational Programme Just Transition and Moravian-Silesian Region.  

\textbf{Competing Interests}
The authors have no relevant financial or non-financial interests to disclose.

\textbf{Data and Code availability}
The data that support the findings of study are openly available in \href{https://github.com/Weliras/RESOURCE-EFFICIENT-VARIATIONAL-QUANTUM-CLASSIFIER}{Github repository} at \href{https://github.com/Weliras/RESOURCE-EFFICIENT-VARIATIONAL-QUANTUM-CLASSIFIER}{https://github.com/Weliras/RESOURCE-EFFICIENT-VARIATIONAL-QUANTUM-CLASSIFIER}.

\textbf{Author contribution.} All authors made equal contributions to the preparation of this paper. All authors wrote the manuscript. PP performed the methodology for the simulations and developed the software, as well as conducted visualization. PL and RK conceptualized the study, performed formal analysis, and supervised the project.








\bibliography{sn-bibliography}

\end{document}